\documentclass[runningheads]{llncs}
\usepackage{graphicx}
\begin{document}

\title{
	On the Tasks and Characteristics of Product Owners: A Case Study in the Oil \& Gas Industry}

\titlerunning{On the Tasks and Characteristics of Product Owners}

\author{Carolin Unger-Windeler \inst{1,2}
	\and
Jil Kl\"under\inst{1}\orcidID{0000-0001-7674-2930}
}
\authorrunning{C. Unger-Windeler}
%
\institute{Leibniz University Hannover, Software Engineering Group, Germany \and
Baker Hughes, a GE Company, Research and Development, Celle, Germany\\
\email{\{carolin.unger-windeler, jil.kluender\} @inf.uni-hannover.de}\\
}
\maketitle              
\begin{abstract}
	Product owners in the Scrum framework -- respectively the on-site customer when applying eXtreme Programming -- have an important role in the development process. They are responsible for the requirements and backlog deciding about the next steps within the development process. However, many companies face the difficulty of defining the tasks and the responsibilities of a product owner on their way towards an agile work environment. 
	
	While literature addresses the tailoring of the product owner's role in general, research does not particularly consider the specifics of this role in the context of a systems development as we find for example in the oil and gas industry. Consequently, the question arises whether there are any differences between these two areas. In order to answer this question, we investigated on the current state of characteristics and tasks of product owners at Baker Hughes, a GE company (BHGE). 
	
	In this position paper, we present initial results based on an online survey with answers of ten active product owners within the technical software department of BHGE. The results indicate that current product owners at BHGE primarily act as a nexus between all ends. While technical tasks are performed scarcely, communication skills seem even more important for product owners in a system development organization. However, to obtain more reliable results additional research in this area is required.

\keywords{Agile Software Development  \and Product Owner \and Systems Engineering \and Systems Development.}
\end{abstract}

\section{Introduction} 
Nowadays, it is a competitive advantage to develop and distribute high-quality software at a high pace \cite{Hohl.2016}. Consequently, software process improvement is a topic many companies have to deal with \cite{Kuhrmann.2018}. For example in the oil and gas industry (but also in other domains such as in the automotive \cite{Hohl.2016}), safety-critical systems are developed that need to be tested thoroughly before they can be rolled out. As a consequence, working software needs to be delivered at an early stage of the system development phase while it still remains flexible and adaptable to changes. Agile software development is a promising possibility to satisfy those needs \cite{Begel.92020079212007,Laanti.2011,Dyba.2008}. However, integrating agile software development practices is often reported as difficult \cite{Boehm.2005}: Introducing agile is not just the introduction of a development method, it is also about changing people by establishing a new mindset \cite{KlunderJilANDSchmittAnnaANDHohlPhilippANDSchneiderKurt.2017}. Becoming agile often goes along with fundamental changes that are facing a lot of barriers \cite{Boehm.2005,Hohl.2016}. 
Regardless of the industry or the company's motivation -- the dilemma is always the same: while the decision of doing agile is made easily, actually becoming agile is not \cite{KlunderJilANDSchmittAnnaANDHohlPhilippANDSchneiderKurt.2017}.  Nevertheless, a lot of companies across all industries strive for it \cite{Klunder.2018}.

This is also the case at \textit{Baker Hughes, a GE Company (BHGE)}. 
BHGE combines capabilities across the full value chain of oil and gas activities -- including the development of digital solutions combining hardware technologies with software products. While hardware engineering has always been one of the company's core businesses, software engineering is relatively new to them.

In daily business, BHGE develops safety-critical systems based on reliable software. To deliver adaptable but high-quality software at an early stage of the system development phase, the management decided to integrate agile development practices. However, BHGE faced difficulties while becoming agile. A previous internal interview study identified the tailoring of the Product Owner (PO) role as the main issue. While becoming agile the main difficulty was -- and still is -- to understand what skills really are required of a PO in this context.

In this contribution, we shed light on the PO at BHGE in order to analyze their tasks and characteristics. To identify the adjusting screws to tailor this role in this context eventually, now tasks and characteristics of active POs are analyzed to asses their current-state and to compare the role of the PO in this context to the PO role described in the literature. In this paper, we present preliminary results based on an online survey conducted at BHGE.


\section{Reference Model of Tasks \& Characteristics of Product Owners}\label{sec:RefMod}
Tailoring the PO role in a system development context is not particularly addressed in current literature. So we looked into the adjacent area of agile software management in large-scale scrum.

\subsection{Related Work: Characteristics of Product Owners}
Pilcher \cite{Pichler.2010} attempted to generate a practical guide that enables new POs to apply agile product management techniques effectively in Scrum. Furthermore, he describes five desirable characteristics of POs, addresses common mistakes when applying this role and suggests a team of POs when it comes to scale this 			role to large projects. The described characteristics are as follows: 

\noindent
\textbf{(1) Communicator \& Negotiator} The PO communicates with and aligns different parties including customers, users, development and engineering, marketing, sales, service, operations, and management.

\noindent
\textbf{(2) Visionary \& Doer} The PO envisions the final product and sees it through to completion. This includes requirements description, closely collaborating with the team, accepting or rejecting work results, and steering the projects by tracking and forecasting its progress.

\noindent
\textbf{(3) Leader \& Team player} The PO is responsible for the product's success, provides guidance for everyone involved and makes tough decisions. He needs to be a team player, rely on close collaboration with other Scrum team members, yet has no formal authority over them.

\noindent
\textbf{(4) Available \& Qualified} Being a PO is usually a full-time job. Project's progress suffers when the PO is overworked. Being adequately qualified usually requires an intimate understanding of the customer and the market.

\noindent
\textbf{(5) Empowered \& Committed}  An empowered PO is essential to bring the product to life. The PO must have the proper decision-making authority -- from finding the right team members to deciding which functionality is delivered as part of the release.

These characteristics are quite high-level and are not sufficient to be checked in the current state analysis at BHGE. To close this gap, the following tasks need to be put in consideration as well.

\subsection{Related Work: Tasks of Product Owners}
Bass \cite{Bass.2015} describes how PO teams scale agile methods to large distributed enterprises. To do so, he identified the following nine PO tasks:

\noindent
\textbf{(1) Intermediary} Act as an intermediary person between all stakeholders.
 
\noindent
\textbf{(2) Traveller} Spend time at client site as well as on all geographical locations of the team to get to know them and disseminate information.

\noindent
\textbf{(3) Communicator} Be available and communicative to all team members to connect teams.

\noindent
\textbf{(4) Techncial Governor} Provide a technical governance framework to project teams in order to ensure usage of common tools and technologies for the project. 

\noindent
\textbf{(5) Release Master} Manage and approve release plans and schedules.

\noindent
\textbf{(6) Prioritizer} Prioritize requirements in the backlog to ensure immediate value to the customer.

\noindent
\textbf{(7) Groom} Gather requirements, translate them into user stories and ensure an evolving backlog.

\noindent
\textbf{(8) Risk Assessor} Evaluate technical complexity and potential shortcomings in the development teams' skills and capabilities. 

\noindent
\textbf{(9) Technical Architect} Design, implement and disseminate a reference architecture between the scrum teams.

Bass \cite{Bass.2015} states that all those tasks should be split on multiple product owners that collaborate in a team. However, in this case study each participating PO is asked about the individual performance regarding each task.

\subsection{Reference Model of Tasks \& Characteristics}
\begin{figure}
	\centering
\includegraphics[scale=0.65]{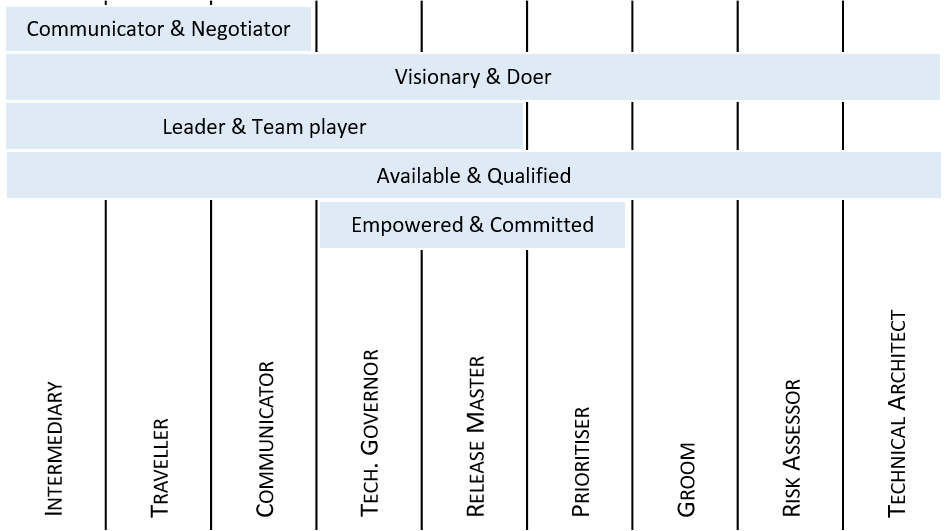}
\caption{Product Owner Characteristics and Tasks} \label{fig1}
\end{figure}

Based on the previously described tasks and characteristics, the reference model shown in Fig. \ref{fig1} is proposed. It is applied to classify the characteristics of the PO at BHGE in terms of the corresponding tasks. As Pilcher \cite{Pichler.2010} and Bass \cite{Bass.2015} do not rate the entities, all of them are considered as equally important. However, some of the characteristics are seen more related to certain tasks (e.g. Communicator \& Negotiator), while others are required on the full range (e.g. Visionary \& Doer).

\section{Product Owner Analysis at BHGE}\label{sec:AnaBHGE}
BHGE develops high-end drilling tools for more than a century. With less accessible oil and gas reservoirs, the drilling process and its tools become increasingly complex. To cope with the increased complexity, digitalization found its way into the drilling technology and the integration of a software development process was needed. 
To deliver high-quality software at an early stage of the system development phase, the technical software group decided to follow agile practices. The Scrum framework was chosen, teams were formed, roles were introduced and all other guidelines adhered.
Though, the integration of an agile software group into a traditional system development environment is challenging at all ends -- a previously conducted internal interview study at BHGE identified the tailoring of the PO role 
as their main issue from a software development perspective.  Understanding the required skills of a PO that fit in this context is still the hardest part. Unfortunately, no literature is particularly discussing this issue. However, related work in the adjacent area of large scale scrum describes required skills of POs in the form of characteristics and tasks. This description is used as a starting point to assess current state of actual POs at BHGE, to distinct the role of a PO in the system developement and to tailor this role in this context eventually.

\subsection{Data Collection}
In order to get an overview of the current tasks and characteristics of POs at BHGE, we conducted an online survey with ten active POs within the technical software department. The participants are located in Germany, the Netherlands, India, USA and do not necessarily work on the same product. They were asked to answer questions that would give some indications about their role in terms of the above mentioned characteristics and tasks. The most considerable part of the survey was structured as a multiple-choice question: \textit{``How would you describe your current role?''} Two possible predefined answers were \textit{``I act as an intermediary person between all stakeholders''} and \textit{``I prioritise the backlog''}. 
To get more detailed information about certain tasks additional questions had a 4-point Likert-scale or were open-ended.

\subsection{Data Analysis}\label{ssec:DA}
\begin{figure}[h]
	\centering
	\includegraphics[scale=0.6]{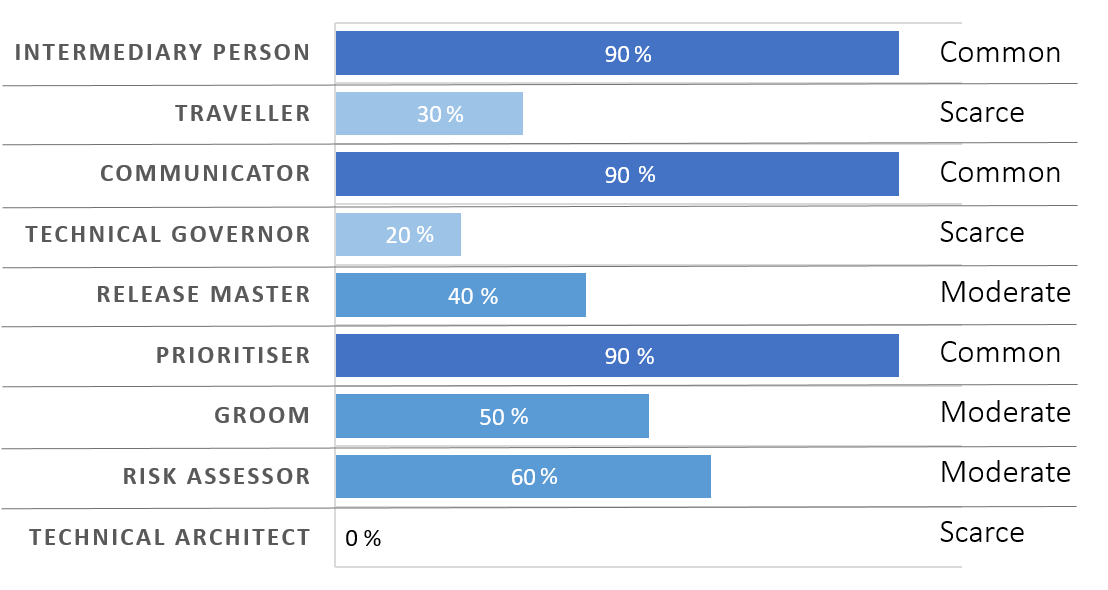}
	\caption{Product Owner Tasks @ BHGE} \label{fig2}
\end{figure}

With the quantitative method of the survey, the statements of the POs regarding their tasks can be summarized in a bar chart. Additionally, based on how many POs performed each task, we divided the tasks into the equidistant intervals summarized in Table \ref{Tab:Tasks}. The results are shown in Fig \ref{fig2}.

\begin{table}[h]
	\centering
	\caption{Division of the tasks}
	\label{Tab:Tasks}
	\begin{tabular}{cc}
		\cline{1-2}
		\textbf{scarce}   &  0--33\% of the POs perform that task  \\
		\textbf{moderate} & 33--66 \% of the POs perform that task \\
		\textbf{common}   & 66--100 \% of the POs perform that task \\
		\cline{1-2} 
	\end{tabular}
\end{table}

\begin{figure}
	\centering
	\includegraphics[width=0.8\textwidth]{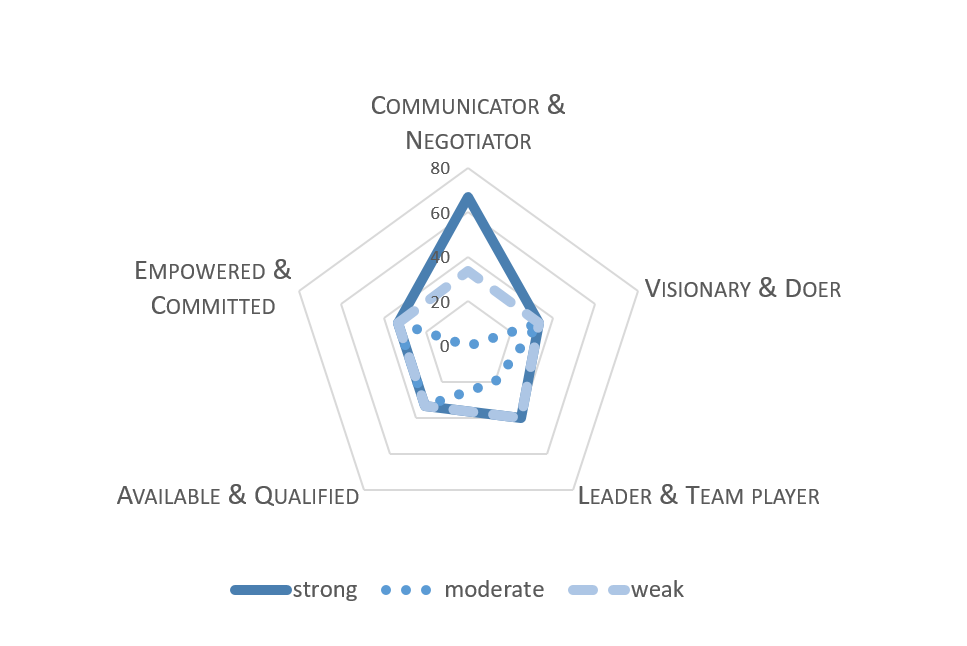}
	\caption{Product Owner Characteristics @ Baker Hughes, a GE Company} \label{fig3}
\end{figure}

\noindent
According to the reference model in Sec. \ref{sec:RefMod} the characteristics of the POs at BHGE can be evaluated as well. Therefore, we again divided the characteristics into the equidistant intervals summarized in Table \ref{Tab:Char}. The results are shown in Fig \ref{fig3}.

\begin{table}[h]
	\centering
	\caption{Division of the characteristics}
	\label{Tab:Char}
	\begin{tabular}{cc}
		\cline{1-2}
		\textbf{weak}     & 0--33\% of the related tasks are performed scarcely\\
		\textbf{moderate} &  33--66 \% of the related tasks are performed moderately\\
		\textbf{strong}   & 66--100 \% of the related tasks are performed commonly\\
		\cline{1-2} 
	\end{tabular}
\end{table}

\section{Results}\label{sec:Res}

\textit{Commonly performed tasks.} The POs at BHGE often act as an intermediary person between all stakeholders and disseminate information across teams, as 9 out of 10 PO would describe their role accordingly. Prioritising the backlog is also something they describe as a common task.

\textit{Moderately performed tasks.} Managing the releases, groom the product backlog, as well as assessing risks are identified as moderately performed tasks as some POs perform those tasks, while others do not.

\textit{Scarcely performed tasks.} The absence of technical decision making is striking. They do not act as a technical governor nor as an architect. Additionally, the POs do not travel much.

\textbf{Considering the tasks, it gets clear that the Product Owners at BHGE are highly communicative but do not make any technical decisions.} 

\textit{Strong Characteristics.} With 2 out of 3 tasks that are mapped to the characteristic of \textit{Communicator \& Negotiator} the POs can be considered as strong communicators and negotiators.
	
\textit{Moderate Characteristics.} All other characteristics are present in a moderate way as most of the mapped tasks are performed moderately. The POs do have a vision, lead the scrum team, are available and qualified, have decisional power and are committed.
	
\textit{Weak Characteristics.} The POs at BHGE do not lack a characteristic according to the reference model.	

\textbf{Overall, POs at BHGE are highly communicative and are empowered to prioritize the backlog according to the stakeholders needs and mostly act as the nexus between all ends. }

The system development organization asks them to communicate and negotiate with all stakeholders -- including the end user of the overall product, leaders of other departments that are involved in the system development as well as the scrum teams. The communicative effort mainly results in the prioritization of the backlog. They are not empowered enough to master the releases as there are too many dependencies to the overall system development plan. POs at BHGE do not make any technical decisions -- neither as a technical governor nor as an architect. A reason is that the currently developed software is replacing legacy systems gradually. Hence, the framework is already set. All architectural decisions are made by a designated architecture team. Another conspicuously scarce performed task is traveling. This is due to the fact that representatives of all involved departments as well as the POs and the end users are co-located.

\section{Conclusion and Future Work}\label{sec:ConFuture}

When BHGE first introduced the role of the PO, they needed to tailor this role to the context of their system development organization. While literature addresses the tailoring of the product owner's role in general, research does not particularly consider the specifics of this role in the context of a system development organization. Consequently, the question arises whether there are any differences between these two areas. 
In order to answer this question, we investigated on the current state of characteristics and tasks of product owners at Baker Hughes, a GE company (BHGE). 
Inital results show that there are differences indeed: being a Product Owner in a traditional, top-down system development organization requires strong communication skills, while technical decisions do not number along the task of a PO as they are made by designated teams. 
According to this findings, the current descriptions of the general PO role is not sufficient 
as some tasks are obsolete while others are missing. 
Also, a recent study of Bass et al.\cite{Bass.2018} identified two more tasks which should be further discussed in this context. 
However, to obtain more reliable results additional research in this area is required.

We hypothesize that a more detailed description of this role will help companies defining the tasks and responsibilities of a product owner on their way towards an agile work environment. 
In future research we will focus on adjusting the tasks and characteristics of PO in a system development context to provide a better understanding of this role. \\


\noindent
\textbf{Acknowledgement.} 
This work was supported by Baker Hughes, a GE Company. 
We commit to upholding the highest ethical standards and complete legal compliance in all we do. Since no data of our employees should be distributed, our data is archived internally for future reference.


%
%
\bibliographystyle{splncs04}
\bibliography{mybib}

\end{document}